\documentclass[fleqn,10pt]{wlscirep}
\title{Disordered Supersolids in the Extended Bose-Hubbard Model}

\usepackage{tabularx}
\usepackage{amsmath}
\usepackage{amssymb}
\usepackage{graphicx}
\usepackage{dcolumn}
\usepackage{bm}
\usepackage{epstopdf}

\author[1]{Fei Lin}
\author[2]{T. A. Maier}
\author[1,*]{V.W. Scarola}
\affil[1]{Department of Physics, Virginia Tech, Blacksburg, Virginia 24061, USA}
\affil[2]{Computational Science and Engineering Division and Center for Nanophase Materials Sciences, Oak Ridge National Laboratory, Oak Ridge, Tennessee 37831, USA}

\affil[*]{scarola@vt.edu}

\begin{abstract}
The extended Bose-Hubbard model captures the essential properties of a wide variety of physical systems including ultracold atoms and molecules in optical lattices, Josephson junction arrays, and certain narrow band superconductors.  It exhibits a rich phase diagram including a supersolid phase where a lattice solid coexists with a superfluid.  We use quantum Monte Carlo to study the supersolid part of the phase diagram of the extended Bose-Hubbard model on the simple cubic lattice.  We add disorder to the extended Bose-Hubbard model and find that the maximum critical temperature for the supersolid phase tends to be suppressed by disorder.  But we also find a narrow parameter window in which the supersolid critical temperature is enhanced by disorder.  Our results show that supersolids survive a moderate amount of spatial disorder and thermal fluctuations in the simple cubic lattice.
\end{abstract}
\begin{document}

\flushbottom
\maketitle

\thispagestyle{empty}

\section*{Introduction}

A superfluid and a solid can, in principle, coexist in the same place at the same time.  This unique state of matter, a supersolid, has attracted numerous research efforts since it was proposed more than 45 years ago in the context of superfluid $^{4}$He \cite{chester:1970,boninsegni:2012}. $^{4}$He experiments suggesting possible observation of a supersolid have remained controversial \cite{kim:2004,kim:2012}.  

Lattice models of supersolids (the extended Bose-Hubbard model \cite{matsubara:1956} in particular) were used to study the critical properties of the supersolid phase of $^{4}$He \cite{matsuda:1970,liu:1972}.  But subsequent work showed that these lattice models capture the essential properties of many other physical systems, including ultracold atoms and molecules in optical lattices \cite{goral:2002,buchler:2003,scarola:2005a,scarola:2006,bloch:2008,baier:2016}, Josephson junction arrays \cite{vanOtterlo:1995,roddick:1995}, and narrow-band superconductors \cite{micnas:1990}.  The latter connection can be made rigorous via a direct mapping between local Cooper pairs and bosons. The supersolid of bosons in this case maps to coexisting superconducting and charge density wave order which has been of interest in a variety of compounds, e.g., $\text{BaBiO}_{3}$ doped with K or Pb \cite{blanton:1993}.  

Experiments with ultracold atoms offer an excellent opportunity to observe supersolids.  These systems are clean and parameters can be tuned to maximize the strength of a supersolid if one is predicted to exist. Different routes to observing  a lattice supersolid have been explored \cite{goral:2002,buchler:2003,scarola:2005a,scarola:2006,bloch:2008,baier:2016}. The key to supersolidity is coexisting order which is, in turn, derived from coexisting diagonal and off-diagonal long-range order in the density matrix.  Along these lines measurements in three dimensions (3D) consistent with coexisting order have been made with ultracold atoms in cavities mediating long-range interactions \cite{baumann:2010,mottl:2012,leonard:2017} and synthetic spin-orbit interactions \cite{li:2017} therefore offering new evidence for supersolids in a controlled and pristine environment.

Work on lattice models suggests that supersolids should be rather delicate and therefore difficult to observe when quantum fluctuations are more pronounced (particularly in lower dimensions).  In two dimensions (2D) it is now known that the lattice supersolid competes with phase separation.  A mean-field argument \cite{sengupta:2005} shows that the formation of domain walls favors phase separation because (for low coordination number) the domain wall intrinsic to the phase separated state gains in kinetic energy.   But on lattices with higher coordination number, e.g., the triangular lattice, quantum Monte Carlo (QMC) calculations show \cite{wessel:2005} that phase separation is suppressed and the lattice supersolid state gains in energy. 

Furthermore, QMC results show that lattice supersolids in 2D are also highly susceptible to disorder.  Results on the square lattice \cite{bernardet:2002} show that spatial disorder destroys the solid itself leaving no chance for the supersolid.  This sensitivity stems from an Imry-Ma-type mechanism \cite{imry:1975,bernardet:2002} implying that the solid is unstable in the presence of arbitrarily weak disorder in less than three dimensions (3D). 

In 3D we expect a solid to be robust against disorder because the Imry-Ma mechanism is avoided \cite{kemburi:2012}.  Furthermore, high coordination numbers have been shown to suppress phase separation on the simple cubic lattice.  QMC results \cite{yamamoto:2009,ohgoe:2012} (in the absence of disorder) report a strong supersolid and no phase separation.  3D lattice models therefore seem to be the best arena to study supersolid behavior.

Study of the extended Bose Hubbard model in 3D has become more pressing because of recent work \cite{baier:2016} that has successfully demonstrated placement of bosonic chromium atoms in a cubic optical lattice.  The atoms have a magnetic dipole moment.  When polarized these moments induce long range interactions.  A theory-experiment comparison \cite{baier:2016}  shows that the extended Bose-Hubbard model quantitatively captures the physics of this system thus paving the way for the possibility of a direct observation of a lattice supersolid as predicted by an extended Bose-Hubbard model.  

\begin{figure}[ht]
\includegraphics[clip,width=150mm]{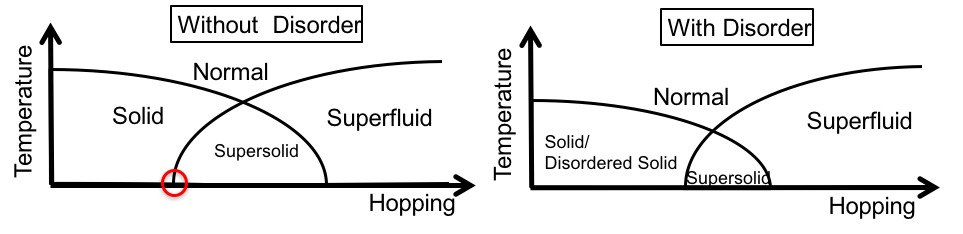}
\caption{{\bf Schematic finite temperature phase diagram.}  Schematic of the phase diagram we find for the extended Bose-Hubbard Model on the simple cubic lattice and at fixed chemical potential with and without disorder.  The left panel shows the absence of order at high temperature (normal phase) while the low temperature phases include the solid, supersolid, and superfluid.  The circle highlights a quantum critical point between the incompressible solid and the compressible supersolid.  In the presence of disorder the supersolid dome shrinks.  The theorem of inclusions \cite{pollet:2009,gurarie:2009} implies that a Griffiths regime must separate the solid from the supersolid phase when disorder is present.  We tentatively assign the intermediate Griffiths regime to a disordered solid phase. In drawing the finite temperature solid phase boundary we note that compressibility is exactly zero only at $T=0$ and decays exponentially to zero at finite temperatures.}
\label{fig:schematic_finite_T_phase_diagram}
\end{figure}

We use QMC to study the stability of the lattice supersolid in the simple cubic lattice.   We study the phase diagram of the extended Bose-Hubbard model.  Our primary results are summarized in Fig.~\ref{fig:schematic_finite_T_phase_diagram} where the phase diagram sketches critical temperature versus lattice hopping energy.  We include disorder in our study to examine the stability of the supersolid.  We find that disorder lowers the critical temperature but the supersolid still survives moderately strong disorder.

 We also find that disorder enhances the supersolid critical temperature \cite{kemburi:2012} within a narrow parameter window.  For low hopping the critical temperature of the solid component of the supersolid remains robust against disorder while the superfluid critical temperature is actually enhanced.  We use systematic finite-size scaling to show that increasing disorder increases the critical temperature of the supersolid.  Our results therefore indicate that supersolids in the extended Bose-Hubbard model are stable (and even enhanced) in the simple cubic lattice for a moderate amount of disorder. Our result is consistent with the disorder effect in dirty superconductors, where one finds that Anderson's Theorem \cite{anderson:1959} is generally observed in weakly interacting regimes, while, for strong interaction, the disorder enhances the superconducting transition temperature due to broadening of the conduction band \cite{kuchinskii:2015}.

\section*{Results}

\subsection*{Model}

We study a tight-binding model of repulsive soft-core bosons hopping in a simple cubic lattice of side lengths $L$ with on-site disorder:
\begin{eqnarray}
H=-t\sum_{\langle i,j\rangle}(a_i^{\dagger}a_j^{\phantom{\dagger}}+\text{H.c.})+\frac{U}{2}\sum_i n_i(n_i-1)
+V\sum_{\langle i,j\rangle}n_in_j-\sum_i \mu_i n_i,
\label{BHmodel}
\end{eqnarray}
where $a_i^{\phantom{\dagger}} (a_i^{\dagger})$ is the boson annihilation (creation) operator at site $i$, $n_i=a_i^{\dagger}a_i^{\phantom{\dagger}}$ is the particle number operator, $t$ is the hopping integral, $U$ is the on-site repulsion, and $V$ is the nearest-neighbor repulsion.  Here $\mu_i=\mu-\epsilon_i$, where $\mu$ is the average chemical potential of the system, and the uniformly distributed random number $\epsilon_i\in [-\Delta, \Delta]$ is the on-site disorder potential.  We use periodic boundary conditions. In contrast to the hard-core boson model, here multiple bosons can occupy the same lattice site in our soft-core boson model, allowing an increase of on-site interaction energy. In the limit of $U\rightarrow\infty$, the soft-core boson model reduces to the hard-core boson model. Hereafter we will use $U$ as the energy unit and set the Boltzmann constant $k_{B}=1$.

For $V=0$ and $\Delta=0$ the model reduces to the well known Bose-Hubbard model \cite{fisher:1989}.   At zero temperature there exist two competing phases, an incompressible Mott insulator at low hopping and a superfluid (SF) at large hopping that spontaneously breaks the continuous $U(1)$ gauge symmetry of the model (the phase invariance of the bosonic operators). At fixed $\mu$ these phases are separated by a quantum critical point at a critical $t$. 

Including a nearest-neighbor repulsion, $V>0$, leads to additional phases.  For large $V$ the bosons tend to sit at every other site to form a charge density wave, a solid (S), which spontaneously breaks the $\mathbb{Z}_{2}$ sublattice symmetry.  When the hopping and interaction terms are comparable a supersolid forms which derives from dual spontaneous symmetry breaking of both the $U(1)$ gauge symmetry and the $\mathbb{Z}_{2}$ sublattice symmetry throughout the entire sample.  The result is simultaneous superfluid and solid order, a supersolid (SS).  To study a regime consistent with spatially decaying interactions and a strong supersolid we choose $zV=U=1$ \cite{sengupta:2005, yamamoto:2009,ohgoe:2012,ohgoe:2012b}, where $z=6$ is the lattice coordination number.  

\begin{figure}[ht]
\includegraphics[clip,width=100mm]{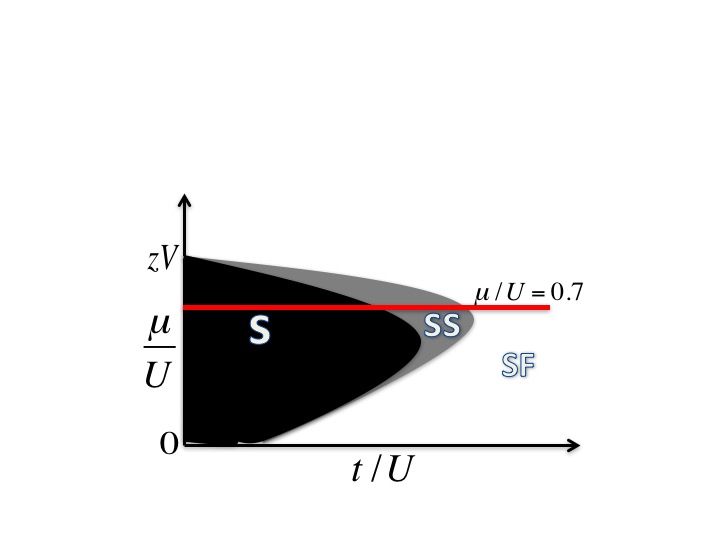}
\caption{{\bf Schematic zero-temperature phase diagram.}  Schematic of the conventional zero temperature mean field phase diagram of Eq.~(\ref{BHmodel}) at density below one with no disorder.  By fixing the chemical potential to be $\mu/U=0.7$ and increasing the hopping from zero we traverse the phase diagram through the solid, supersolid, and superfluid phases, respectively \cite{vanOtterlo:1995,scarola:2006}.}
\label{mft_diagram}
\end{figure}

Mean-field analyses of the disorder-free extended Bose-Hubbard model \cite{vanOtterlo:1995,scarola:2006,kemburi:2012} show that the supersolid sits between the solid and the superfluid in the phase diagram.  Fig.~\ref{mft_diagram} shows the zero temperature mean field phase diagram in the dilute (low $\mu$) regime with $\Delta=0$.  In the following, we select a specific average chemical potential, $\mu=0.7$, unless otherwise stated.  The horizontal line indicates that increasing $t$ while keeping $\mu=0.7$ allows us to transverse three of the phases discussed so far, i.e., S$\rightarrow$SS$\rightarrow$SF.  This choice also keeps the density at or below one.  By adding disorder to phases lying along the horizontal line, we can obtain other intriguing phases, such as, the Bose glass (BG) \cite{fisher:1989}.  We also identify a compressible regime which maintains the character of a solid (long range order in the density).  We call this regime a disordered solid (DS) phase and assume it plays the role of the Griffiths phase which must \cite{pollet:2009,gurarie:2009} lie between the incompressible solid and compressible supersolid phases.  This assignment follows from a low $t$ mapping to disordered phases of the Spin-1 Blume-Capel model (See Supplemental Material).  

To focus primarily on the interplay of the supersolid with disorder we focus our study on weak disorder.  We exclude higher disorder here because proper order parameters for finite size scaling analyses  to determine the BG and DS phase boundaries at finite temperatures are unknown. Since there has been some recent progress in constructively identifying the BG phase using local condensate fraction \cite{meldgin:2016} at finite temperatures we think future studies of strong disorder phases should be able map out the phase boundaries.

\subsection*{Definition of Order Parameters}

Each of the states discussed as low temperature phases of Eq.~(\ref{BHmodel}) correspond to unique combinations of order parameters.  At high temperatures the normal phase (N) is defined by the absence of order (either local or non-local).  Whereas low temperature regimes tend to show order in either the diagonal or off-diagonal parts of the single-particle density matrix (or both as in the supersolid phase).  This section lists the phases we find and the corresponding order parameters.  

Solid order is defined by long-range oscillations in the density-density correlation function (diagonal long-range order in the density matrix) or, equivalently, peaks in the static structure factor at wavevector, ${\bf Q}$:
\begin{equation}
S_{Q}\equiv N_{s}^{-2}\sum_{j,k}e^{i\mathbf{Q}\cdot (\mathbf{r}_j-\mathbf{r}_k)}\langle n_jn_k\rangle,
\nonumber
\end{equation}
that indicate a spontaneous breaking of the sublattice symmetry.  $N_{s}=L^{3}$ is the number of sites.  For the large values of $V$ considered here an oscillation of the density between sublattices is favored, i.e., ${\bf Q}=(\pi,\pi,\pi)$ on the simple cubic lattice.  

The solid phase we discuss here is incompressible.  The compressibility defines 
how easily the particle number fluctuates in the system, and is given by:
\begin{equation}
\kappa=\frac{\partial \rho}{\partial \mu}=\frac{N_{s}}{T} \left [  \langle \hat{N}^{2} \rangle- \langle \hat{N}\rangle^2 \right ].
\label{kappatotal}
\end{equation}
where $T$ is the temperature and the average particle density is given by:
\begin{equation}
\rho\equiv \langle \hat{N} \rangle
\nonumber
\end{equation}
with:
\begin{equation}
 \hat{N}\equiv N_{s}^{-1}\sum_{i} n_{i}.
 \nonumber
\end{equation}
The last equality in Eq.~(\ref{kappatotal}) shows that the compressibility is intrinsically non-local because it relates to density fluctuations across the entire system, $ \langle \hat{N}^{2} \rangle$. 

The superfluid density describes the system's response to external perturbations, such as translation or rotation.  It is characterized by off-diagonal long-range order in the density matrix even in the presence of interactions.  In the path-integral QMC formalism the superfluid density is given by \cite{pollock:1987}:
\begin{equation}
\rho_s=\frac{T\langle W^2\rangle}{3t\rho},
\nonumber
\end{equation}
where the squared winding number is $W^2=W_x^2+W_y^2+W_z^2$ and $W_i$ is the winding number in the $i$th direction with $i=x$, $y$, or $z$.  We find that the above order parameters adequately characterize the low-temperature phases of Eq.~(\ref{BHmodel}) at weak disorder.  

\begin{figure}[ht]
\includegraphics[clip]{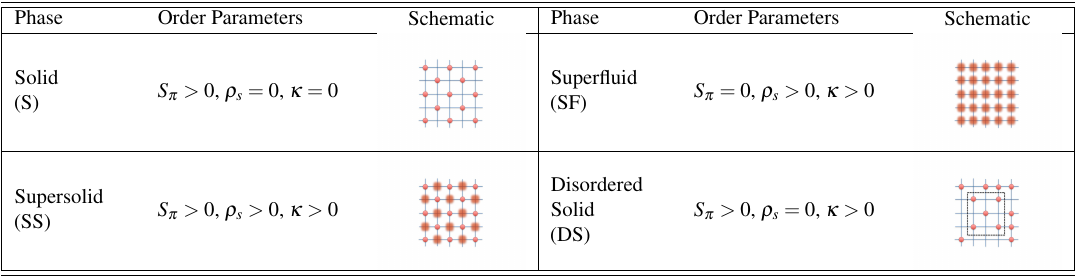}
\caption{{\bf Schematics of Order Parameters.} Schematics of low temperature phases of Eq.~(\ref{BHmodel}) and related order parameters.  A solid (blurred) sphere 
represents a localized (delocalized) particle.  Delocalized particles represent non-zero superfluid density. The black square in the DS schematic shows the short-range checker-board pattern; such a pattern disappears outside the box, but the box is repeated throughout the sample. The normal and Bose-glass phases are not shown but have $S_{\pi}=0$, $\rho_s=0$, $\kappa>0$ (no long-range order), and occur at high and low temperatures, respectively.}
 \label{table:orderparameter}
\end{figure}

Figure~\ref{table:orderparameter} summarizes the order parameters and the phases we discuss.  As we vary $T$, $t$, and $\Delta$, we find the following phases:  normal, solid, superfluid, and supersolid.  We also find evidence for a disordered solid in finite-size simulations.  The absence of order at high $T$ signals the normal phase.  The system forms a  $\rho=1/2$ solid when it has long-range diagonal order, $S_{\pi}>0$, while maintaining incompressibility, $\kappa=0$.   Superfluid order is described by $\rho_s>0$ and $\kappa>0$.  To obtain supersolid order, the system needs to have coexisting solid and superfluid orders, i.e., $S_{\pi}>0$, $\rho_s>0$, and $\kappa>0$.  The disordered solid arises in the presence of disorder.  Defects lead to domains with gapless edges that leave the system compressible, i.e., $S_{\pi}>0$ but  with $\kappa>0$.  The Bose-glass phase occurs for large disorder strengths.  It has only local superfluid order (no off-diagonal long-range order).  It is compressible but exists only at low $T$.

\subsection*{Quantum Monte Carlo Evaluation of Order Parameters}

This section summarizes QMC calculations of order parameters as a function of parameters in Eq.~(\ref{BHmodel}).  Parameter sweeps are used to qualitatively identify regions of the phase diagram with (and without) disorder.  These parameter sweeps are then used to find phase boundaries using finite-size scaling.

\begin{figure*}[ht]
\includegraphics[clip]{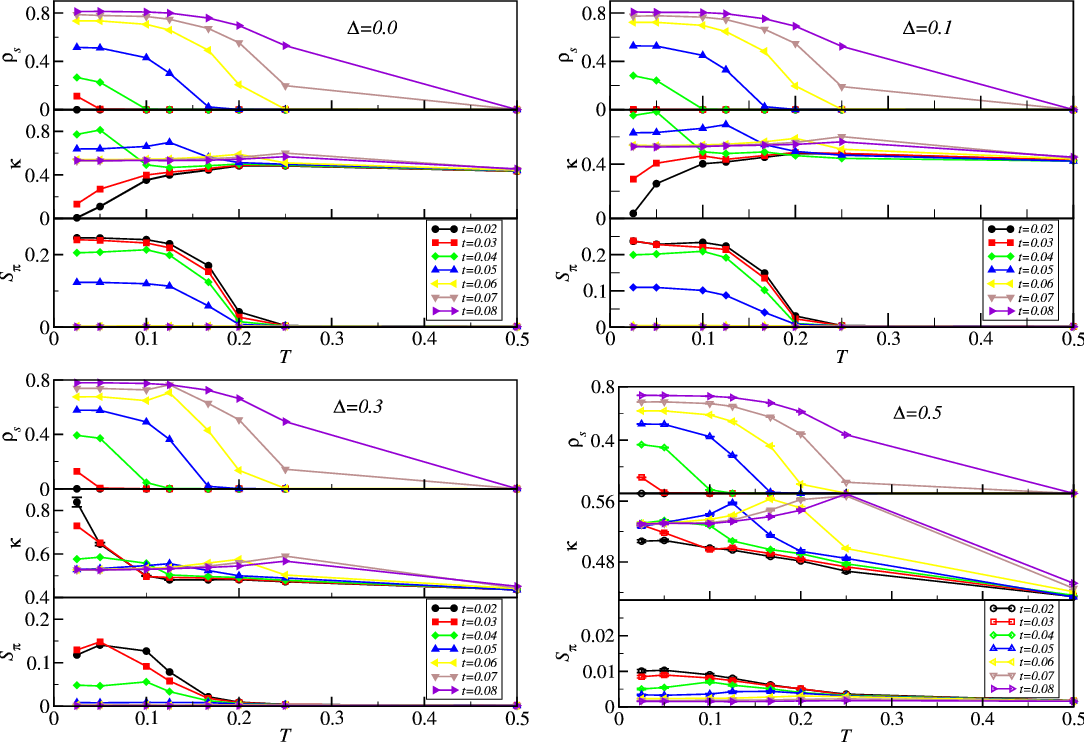}
\caption{{\bf Order parameters versus temperature.} Order parameters $\rho_s$, $\kappa$, and $S_{\pi}$ computed as functions of temperature $T$ and hopping $t$ for various disorder strengths using quantum Monte Carlo on Eq.~(\ref{BHmodel}) for $L=10$.  Here and in the following the error bars are smaller than the symbol size unless depicted otherwise and the lines are guides to the eye.  We set $\mu=0.7$ and $V=1/z$ here. Parameters $t$ and $\Delta$ are shown in the figure. }
\label{fig:rough_phase_scan}
\end{figure*}

To qualitatively locate phases on the $T$ vs. $t$ phase diagram of Eq.~(\ref{BHmodel}) we scan $t$ as well as disorder $\Delta$.  We choose four disorder strengths: 
$\Delta=0.0$, 0.1, 0.3, and 0.5.  To obtain temperature dependence we also sample the following set of temperatures: $T=0.025, 0.05, 0.1, 0.125, 0.167, 0.2, 0.25$, and $0.5$.  We first do these simulations at $L=10$ for our qualitative estimate.  
Fig.~\ref{fig:rough_phase_scan} plots the order parameters of the model as a function of temperature for several different $t$.  The top panels plot the superfluid density.  We can compare all four top panels to see that for large $t$ the disorder does not suppress the superfluid density much.  We can understand this effect using the mapping to the attractive Fermi-Hubbard model (See Supplementary Material) where the superfluid corresponds to an $s$-wave superconductor.  The robustness of the superfluid found here then follows from the Anderson's theorem \cite{anderson:1959} for the robustness of $s$-wave superconductivity to disorder.    

The middle panels in Fig.~\ref{fig:rough_phase_scan} plot the compressibility.  Here we see that the finite size of the system keeps $\kappa>0$ for all but the lowest $t$ and $t$ with $\Delta=0$.  Using finite-size scaling we find that the solid phase is incompressible in the thermodynamic limit.  

The bottom panels in Fig.~\ref{fig:rough_phase_scan} plot the structure factor.  Here we see that at large $t$ and/or $T$, the structure factor vanishes.  This indicates that we have either the superfluid or normal phase.  But for low $t$ and low $T$ the structure factor increases to reveal a supersolid and, for very low $t$, a pure solid.  As disorder is increased to $\Delta=0.1$ the pure solid gives way to what appears to be a compressible phase.  Here the distinct $T$ dependence of the compressibility indicates a distinction from the pure solid.  We tentatively assign this regime to be the disordered solid phase.    

At $\Delta=0.1$, we find that the supersolid phase at $t=0.03$ is destroyed. The system has $\rho_s=0$, $\kappa>0$ and $S_{\pi}>0$, satisfying the definition of the disordered solid phase. By increasing $t$ to $t=0.04$, we recover the supersolid phase, which persists until $t=0.06$.  For larger $t$ the system enters the superfluid phase.  $\Delta=0.3$ shows a similar set of transitions. However, the critical temperature for Bose-glass to normal phase transition is undetermined in our study, since we have not found a suitable scaling relation to describe the transition.

For $\Delta=0.5$ and small $t$ ($t=0.02$), we have $\kappa>0$, $\rho_s=0$, and $S_{\pi}\sim 0$, which is the Bose-glass phase at low $T$. Our result is consistent with the existence of Bose-glass phase predicted by theorem of inclusions \cite{pollet:2009b}. As we increase the hopping to $t=0.03$, the system turns into the superfluid at low temperatures. The superfluid phase persists as we further increase $t$ values. 

Note that the above rough determination of phase boundaries for $L=10$ will change with system size. The critical points deduced from Fig.~\ref{fig:rough_phase_scan} are only approximate. Precise determination of critical points can be achieved through finite-size scalings to be discussed below. However, from the numerical simulations at $L=10$, we already see the rich phase diagram contained in the disordered extended Bose-Hubbard model. Numerical simulations at $L=10$ also serve as a rough guide to phase transitions, which will suggest parameters for a precise finite-sizing scaling analysis.

\subsection*{Finite-Size Scaling}

To map out the finite temperature phase boundaries, we used QMC data to carry out finite-size scaling analyses for the order parameters.  We found two distinct universality classes governing transitions: Ising and  3D XY.  The Ising universality class applies to the long-range charge order/disorder transition while the 3D XY universality class applies to superfluid/non-superfluid transitions.  In this section we discuss the methods we used to identify the transition points using finite-size scaling relations.  

\begin{figure*}[ht]
\includegraphics[clip]{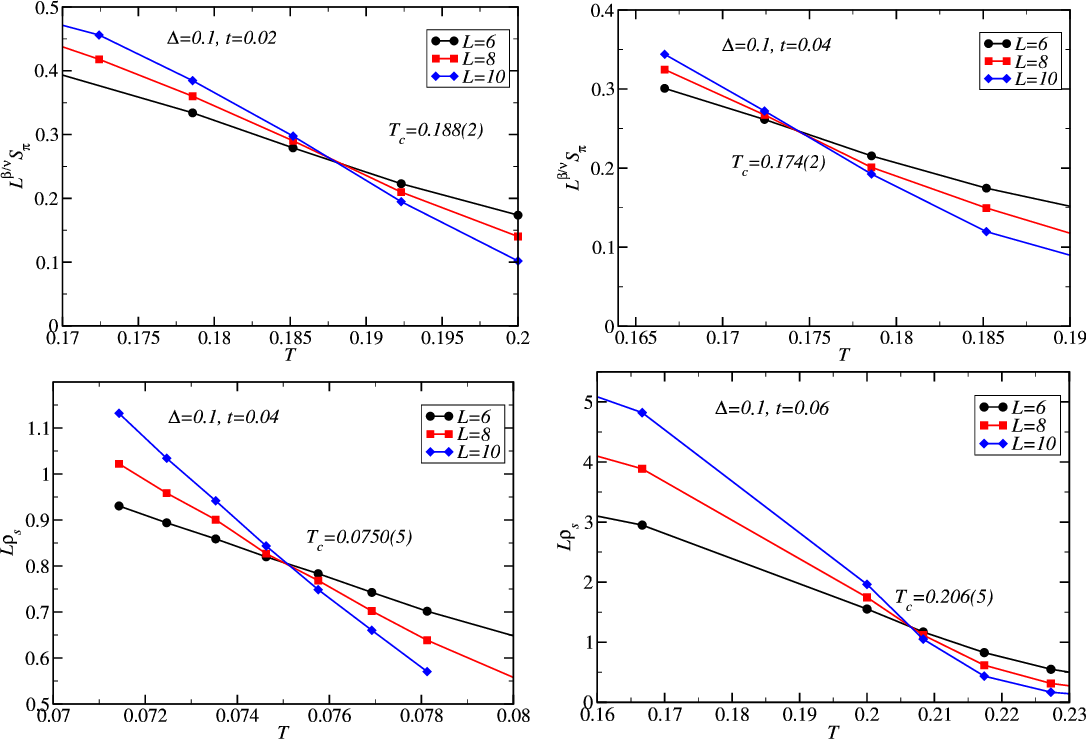}
\caption{{\bf Finite-size scaling for thermal phase transitions.} Example rescaled order parameters plotted as a function of temperature computed using quantum Monte Carlo on Eq.~(\ref{BHmodel}).  The crossings of the rescaled structure factor and superfluid density allow determination of the critical temperatures for order parameters $S_{\pi}$ and $\rho_s$, respectively. The error bars are estimates in the uncertainty of the crossing points. Here we set $\mu=0.7$ and $V=1/z$. Parameters $t$, $\Delta$, and $L$ are shown in the figure.}
\label{scalinganalysis}
\end{figure*}

Since the long-range charge order to disorder transition belongs to the Ising universality class, the structure factor obeys the following scaling relation \cite{privman:1984}:
\begin{equation}
S_{\pi}=L^{-\beta/\nu}\tilde{S}_{\pi}(a_1L^{1/\nu}\tilde{t}),
\label{SpiScaling}
\end{equation}
where $\tilde{t}=(T-T_c)/T_c$ is the reduced temperature that measures the dimensionless distance from $T$ to the critical temperature $T_c$, $\beta=0.3265(3)$, $\nu=0.6301(4)$, 
$a_1$ is a non-universal metric factor, and $\tilde{S}$ is a scaling function. From Eq. (\ref{SpiScaling}) we see that if we plot $L^{\beta/\nu}S_{\pi}$ vs. $T$ for different lattice sizes, different curves will intersect at $T=T_c$.  Two example scaling figures are shown in the upper two panels of Fig.~\ref{scalinganalysis} for $\Delta=0.1$.

On the other hand, the 
superfluid to non-superfluid transition belongs to the 3D XY universality class, and the superfluid density scaling satisfies the following scaling relation \cite{cha:1991}:
\begin{equation}
\rho_s=L^{-(d-2)}\tilde{\rho}_s(a_2L^{1/\nu}\tilde{t}),
\label{rhoScaling}
\end{equation}
where $\tilde{\rho}_s$ is a scaling function, $d=3$ is system dimension, and $a_2$ is a non-universal metric factor.

In 3D, we can plot $L\rho_s$ vs. $T$ for different lattice sizes. Different curves again intersect at $T=T_c$ for the transition.  The lower two panels of Fig.~\ref{scalinganalysis} show example finite-size scaling analyses of $\rho_s$ for disorder strength $\Delta=0.1$.  We have checked that the $L=6-10$ data are sufficient to give accurate critical points by including larger system sizes ($L\leq20$) for select parameters.  

\begin{figure}[ht]
\includegraphics[clip,width=90mm]{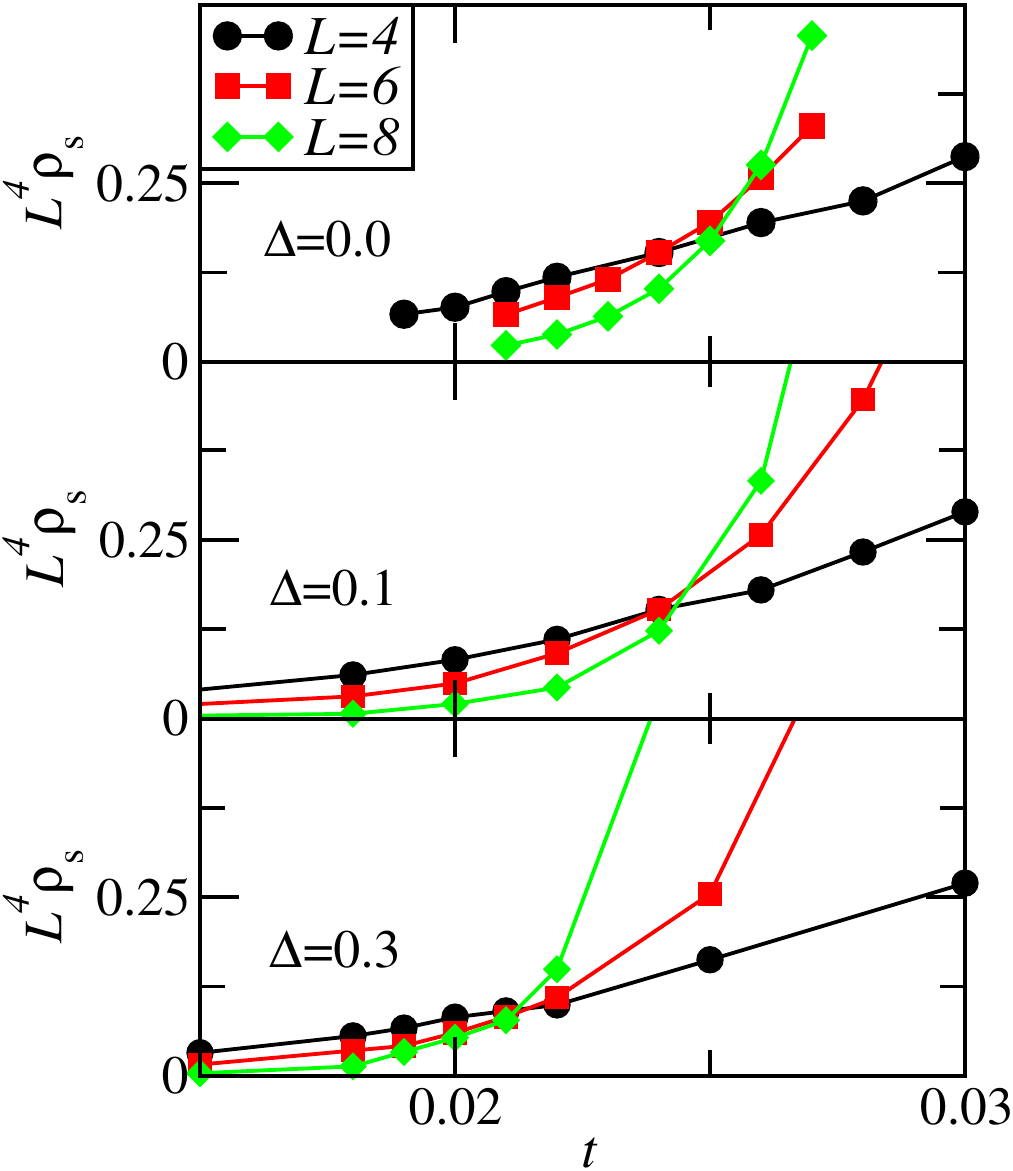}
\caption{ {\bf Finite-size scaling for quantum phase transitions.}  Finite-size scaling of the stiffness as a function of the hopping for three different system sizes, $L=4,6,$ and $8$, for $\mu=0.7$ and $V=1/z$. The top, middle, and bottom panels correspond to disorders $\Delta=0, 0.1$, and $0.3$, respectively, where we obtain critical points, $t_c=0.025(1), 0.0245(5),$ and $0.021(1)$ from the crossing of all three curves.  The error bars are estimates in the uncertainty of the crossing points. 
  }
\label{fig:combine_T0_rhos_vs_t_scaling}
\end{figure}

We also perform scaling analysis to locate the quantum critical point 
$t_c$ for superfluid density as we vary $t$. The superfluid 
density satisfies the following scaling relation 
\cite{privman:1984, lin:2011a}:
\begin{equation}
\rho_s=L^{\alpha}f(aL^{1/\nu}\delta, L^{-z}/T),
\end{equation}
where $\alpha=2-d-z$, $\delta=t-t_c$ measures the 
distance to the critical point, $z$ is the dynamical exponent, which 
is predicted to be $z=d$ \cite{fisher:1990}, $a$ is a non-universal metric number, and the function $f$ is universal. For our cubic lattice, we 
have $\alpha=-4$.  Fig.~\ref{fig:combine_T0_rhos_vs_t_scaling} shows results from QMC simulations, where we keep $L^{-3}/T=0.03125$. Hence, for 
$L=4, 6,$ and $8$, simulations are carried out at $T^{-1}=2, 6.75, 16$, respectively, for various $t$ values around the critical point.  Using these scaling relations we are able to locate phase transition lines to construct a phase diagram for the supersolid. Note that it is also possible to determine the critical temperatures using data collapse technique \cite{binder:1980} with the above scaling relations. However, one would need to perform more QMC simulations with a much finer temperature grid for various lattice sizes. 

\subsection*{Phase Diagrams}

This section culminates the results and methods presented in previous sections to construct QMC phase diagrams of Eq.~(\ref{BHmodel}).  Finite-size scaling of the superfluid stiffness and the structure factor are used to find finite temperature critical points for the solid, supersolid, and superfluid phases.   Finite-size scaling is also used to get the quantum critical points as a function of $t$.  We find that disorder tends to suppress the supersolid critical temperature over much (but not all) of the phase diagram.  Our central finding is that the supersolid is present at intermediate hopping even in the presence of disorder.   

\begin{figure}[ht]
\includegraphics[clip,width=90mm]{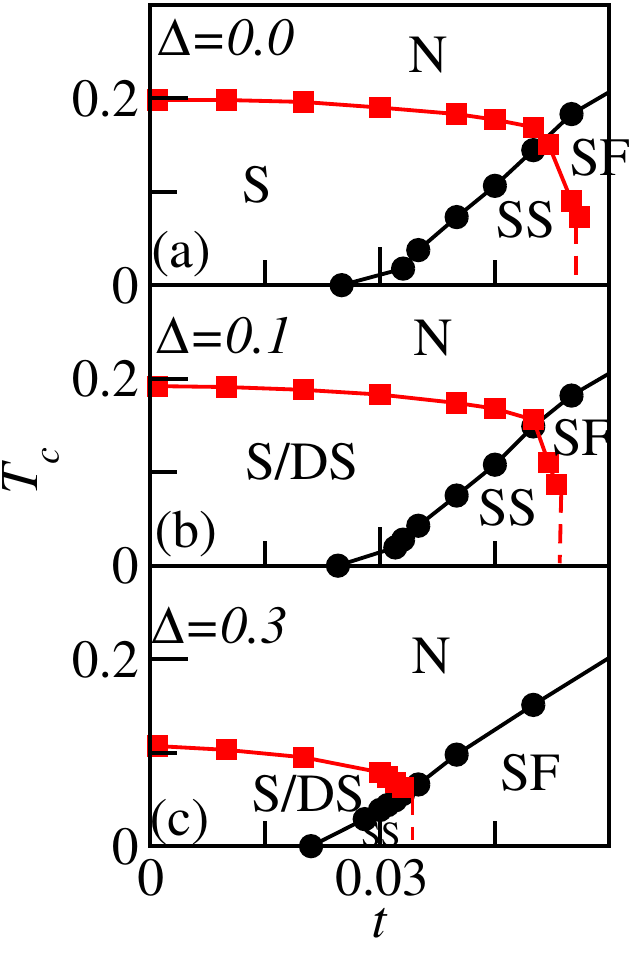}
\caption{{\bf QMC phase diagrams.} Critical temperatures of the extended Bose Hubbard model, Eq.~(\ref{BHmodel}), computed using quantum Monte Carlo for different disorder strengths.  Squares result from the $S_{\pi}$ scaling analysis [Eq.~(\ref{SpiScaling})] and circles are from the $\rho_s$ scaling analysis  [Eq.~(\ref{rhoScaling})]. The dashed lines indicate expected phase boundaries inferred from finite-size estimates that could not be obtained by our method in the thermodynamic limit. }
\label{fig:phase_diagram_allD}
\end{figure}

The $\Delta=0$ panel in Fig.~\ref{fig:phase_diagram_allD} plots the phase diagram of Eq.~(\ref{BHmodel}) in the absence of disorder as determined by QMC.  Here squares and circles plot the critical temperature determined by finite-size scaling of the structure factor and the stiffness, respectively.  We see that the solid and superfluid dominate at small and large hopping, respectively.  The supersolid is found at intermediate hoppings.  

The vertical dashed lines in Fig.~\ref{fig:phase_diagram_allD} indicate an expected phase boundary.  Our conclusions here are based on finite-size data without extrapolation.  For example, for $\Delta=0$ increasing system size drives the critical temperature to zero for $t\gtrsim 0.0525$.  Here we were not able to resolve the critical temperature uniquely given our method because the phase boundary is nearly vertical here.  

The remaining panels in Fig.~\ref{fig:phase_diagram_allD} plot the same as the top panel but in the presence of disorder.  We find that increasing spatial disorder tends to lower the maximum critical temperature of the solid phase.  Here the $T_c$ of the solid order tends to be more sensitive to disorder than the superfluid.  It is therefore the lowering of $T_c$ of the solid that suppresses the supersolid behavior.

Griffiths effects should be particularly important in the thermodynamic limit near phase boundaries separating incompressible and compressible phases \cite{pollet:2009,gurarie:2009}.  The solid and supersolid are incompressible and compressible, respectively.  Our phase diagrams omit the quantum Griffiths phase which, according to the theorem of inclusions \cite{pollet:2009}, must separate these two phases.  We tentatively assign the intermediate quantum Griffiths regime to be a disordered solid (in analogy to the Bose-Glass in the ordinary Bose-Hubbard model \cite{pollet:2009,gurarie:2009}) based on our preliminary finite-size results (Fig.~\ref{fig:rough_phase_scan}).   We have not been able to use finite-size scaling to identify the $T_c$ for the disordered solid.  We therefore label the solid phase in the presence of disorder in Fig.~\ref{fig:phase_diagram_allD} as S/DS to allow for the disordered solid phase between the solid and supersolid phases.

\subsection*{Disorder Enhanced Supersolids}

The addition of disorder can, counterintuitively, enhance supersolidity in a narrow parameter window of the phase diagram.  We first consider the impact of disorder on the solid component of the supersolid.  For large hopping, $t\gtrsim0.035$,  the disorder suppresses the $T_c$ of the solid because the disorder destroys translational invariance required by the solid.   But for low $t$, the solid is more robust and weak disorder does not significantly impact $T_c$ of the solid.  There is therefore a narrow regime (we find it to be near $t\approx0.03$) where the $T_c$ of the solid component of the supersolid is not significantly impacted by disorder.

The superfluid component of the supersolid, on the other hand, can be increased by disorder.  Previous work looking at the ordinary Bose-Hubbard model  ($V=0$) found that the $T_c$ of the superfluid can be increased by disorder \cite{krauth:1991,gurarie:2009, lin:2011a,kemburi:2012}.  The mechanism required disorder to create pathways for the superfluid to percolate across the entire sample.  The pathways enlarged the phase space for superfluidity, and therefore $T_c$. 

\begin{figure}[ht]
\includegraphics[clip,width=100mm]{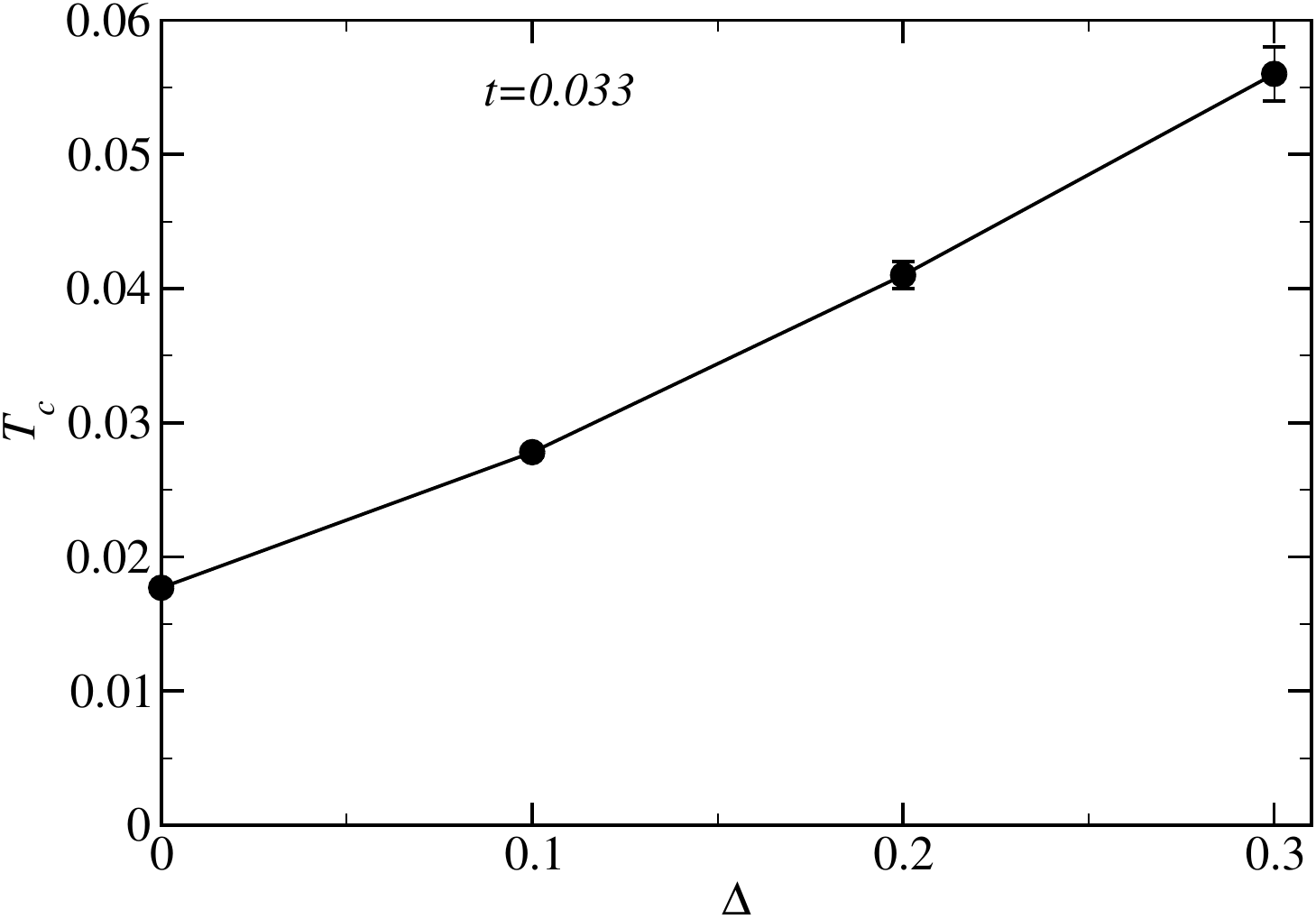}
\caption{{\bf Supersolid critical temperature versus disorder.} Circles plot the critical temperature of the supersolid phase as a function of disorder strength computed using quantum Monte Carlo on Eq.~(\ref{BHmodel}) along with finite-size scaling.  These results show that disorder can, for certain hopping strengths, enhance the critical temperature for the supersolid phase.  For disorder strengths above $\Delta=0.3$, the critical temperature quickly drops to zero.  }
\label{SSbyDisorder}
\end{figure}

The combined effects of a stable solid with enhanced superfluidity leads to an enhanced $T_c$ for the supersolid with disorder.  To see this in QMC we use finite-size scaling to show that disorder can increase the critical temperature of the supersolid phase in the thermodynamic limit in a narrow parameter window.  We extract critical temperatures for the supersolid for various disorder strengths ($\Delta=0, 0.1, 0.2,$ and $0.3$). Finite-size scaling is performed for lattice sizes $L=6, 8, 10$.  Fig.~\ref{SSbyDisorder} shows an enhancement of the critical temperature for the supersolid phase from $T_c\sim 0.02$ to $T_c\sim 0.06$ as we increase the disorder strength from $\Delta=0.0$ to $\Delta=0.3$ at $t=0.033$.  An approximate 3-fold increase of the supersolid critical temperature is achieved by increasing disorder in a narrow window of $t$.  Fig.~\ref{SSbyDisorder} is consistent with previous results \cite{kemburi:2012} but carries the calculation into the thermodynamic limit with an explicit calculation of $T_c$.  $T_{c}$ drops quickly for larger disorder strengths.

The disorder enhanced supersolid can also be understood in a mean-field percolation picture \cite{kemburi:2012}.  Consider the pure solid phase near the solid-supersolid phase boundary in the absence of disorder.  The gap in the solid phase prevents density fluctuations and therefore suppresses inter-site tunneling needed for concomitant superfluidity.  The addition of site disorder allows tunneling between sites with sufficiently strong disorder.  If the collection of bonds allowing tunneling percolates across the sample, then a superfluid forms.  In this way the superfluid has been found to be triggered by the addition of disorder in Bose-Hubbard models  \cite{krauth:1991,sheshadri:1995,dang:2009,pollet:2009,gurarie:2009}.  But here the background solid remains intact leading to a supersolid that has been triggered by the addition of disorder.

\section*{Discussion}

We have used quantum Monte Carlo to study the extended Bose-Hubbard model with disorder on the simple cubic lattice.  We have computed the finite temperature phase diagram at fixed chemical potential.   We find that disorder lowers the maximum critical temperature of the supersolid.  But our results show that disorder and thermal fluctuations still allow the supersolid phase, in contrast to lower dimensions where disorder suppresses the supersolid \cite{imry:1975}.   We have also found that in a narrow parameter regime, the critical temperature of the supersolid is actually enhanced by disorder where the disorder opens percolating pathways to strengthen superfluidity.  Overall, our results show that in 3D the supersolid is more robust than in lower dimensions \cite{sengupta:2005}.

\section*{Methods}

We solve Eq.~(\ref{BHmodel}) using a numerically exact QMC method: the Stochastic Series Expansion representation with directed loop updates \cite{sandvik:1999,syljuasen:2002}.  Various physical quantities, either diagonal or off-diagonal, can be calculated according to the path integral formulation of the QMC simulations.  Our results are converged with respect to truncation of the boson number, the number of QMC steps, and the number of disorder profiles.  Our estimates of order parameters are therefore exact to within Monte Carlo error.  We have also checked that our implementation of the Stochastic Series Expansion algorithm produces the same results as the ALPS implementation \cite{bauer:2011}.

Disorder averaging is a key part of the numerical procedure.  We perform several runs over distinct disorder profiles to ensure proper averaging.  To ensure convergent disorder averages, we typically run 1000 QMC simulations with different disorder realizations for each set of parameters. We then plot histograms for the resulting measurement of various physical quantities.  

We find three types of distributions in our disorder averaging.  The most common distribution is a single Gaussian peak without 
any ``fat'' tails in the distribution curve.  This type of distribution signifies a unique phase for the parameter set.  A Gaussian distribution offers fast convergence with respect to the number of disorder realizations.  

We also find double-peaked Gaussian distributions at low $T$ and large systems, $L\geq10$.  Our QMC simulations usually end up in one of the two phases depending on the initial configuration.  In this case, numerical data are sorted according to the two phases and separate averages need to be done, one for each phase.  We choose the phase with the lowest free energy. It is worth noting that two-peak distribution does not necessarily imply  the coexistence of two phases. Instead we believe that the two-peak structure is due to trapping in a free energy local minimum (the small peak in the distribution, usually less than $5\%$ of the disorder samples).  Updates are then unable to find a path to the free energy global minimum (the large peak in the distribution). We have checked our calculations against ALPS code \cite{bauer:2011}, and found that ALPS exhibits the same trapping.   

The third type of distribution is a single Gaussian peak but with a ``fat'' tail \cite{lin:2011a}.  This happens in the Bose-glass phase, where our order parameters do not assume a definite value.  In this case physical quantities will have a slow convergence rate with respect to disorder configurations \cite{lin:2011a}.

The datasets generated during and/or analyzed during the current study are available from the corresponding author upon reasonable request.

\bibliography{foo.bbl}

\section*{Acknowledgements}

V.W.S. acknowledges support from AFOSR (FA9550-15-1-0445) and ARO (W911NF-16-1-0182). TAM acknowledges support from the Center for Nanophase Materials Sciences, which is a DOE Office of Science User Facility.

\section*{Author contributions statement}

F.L. performed numerical simulations.  All authors analyzed the results and reviewed the manuscript. 

\section*{Additional information}

The authors declare no competing financial interests.

\end{document}